\documentclass[]{article}

\usepackage[T1]{fontenc}
\usepackage[utf8]{inputenc}
\usepackage{graphicx}
\usepackage{multirow}
\usepackage{RR}
\pdfoutput=1

\usepackage{bibentry}
\nobibliography*

\usepackage{tabularx}

\usepackage[table]{xcolor}

\usepackage{eso-pic}
\makeatletter
   \AddToShipoutPicture{%
 }
\makeatother

\RRNo{7952}
\RRdate{05/2012}
\RRauthor{Benoit Baudry \and Martin Monperrus}
\authorhead{Grimm \& Louarn \& others}
\RRtitle{Towards Ecology Inspired Software Engineering}
\RRetitle{Towards Ecology Inspired Software Engineering}

\titlehead{Towards Ecology-Inspired Software Engineering}
\RRresume{
Les écosystèmes sont des systèmes complexes et dynamiques.
Au cours de l'évolution, ils ont développé des capacités avancées pour fournir des fonctions stables, et ce malgré des changements constants dans l'environnement.
Dans ce papier, nous discutons l'hypothèse que les lois dirigeant l'organisation et le développement des écosystèmes sont une source d'inspiration riche pour l'architecture et la construction des logiciels.
}
\RRabstract{
Ecosystems are complex and dynamic systems. Over billions of years, they have developed advanced capabilities to provide stable functions, despite changes in their environment. In this paper, we argue that the laws of organization and development of ecosystems provide a solid and rich source of inspiration to lay the foundations for novel software construction paradigms that provide stability as much as openness.
}
\RRmotcle{génie logiciel, écologie, étude empirique}
\RRkeyword{software engineering, ecology, empirical study}
\RRprojets{Triskell and Adam}
\RCRennes 

\begin{document}

\author{
\begin{minipage}{5cm}\centering Benoit Baudry\\ INRIA \end{minipage}
\begin{minipage}{5cm}\centering Martin Monperrus\\ University of Lille \& INRIA\end{minipage}
}
\date{}

\makeRR


\section{Introduction}

Current expectations about software intensive systems are extremely high: they should provide a large number of rich services,  sense and act on the physical world, and we expect them to be open in the sense of being able to accept  unforeseen components and features. Meanwhile, we expect those systems to be reliable, available and secure. A major challenge  lies in the complexity required to build open systems (first expectation), which hampers all kinds of stability (second expectation). 

Ecosystems are \emph{``made up of all the organisms in a given area interacting with the physical environment''} \cite{Odum1969}. They are complex and dynamic systems. Over billions of years, they have developed advanced capabilities to provide stable functions, despite changes in their environment (e.g. a drought). 
We think that the laws of organization and development of ecosystems provide a solid and rich source of inspiration to lay the foundations for novel software construction paradigms that provide stability as much as openness. 

 Ecology is the science which studies the structure and dynamics of ecosystems.
In other words, the literature from Ecology addresses the comprehension of systems that seem comparable to today's software systems.
The scientific intuition behind this paper is that a number of concepts and methods from Ecology provide a solid and rich source of inspiration for inventing new software construction paradigms. 

Several lines in software engineering claim an inspiration from ecosystems. Bernardo A. Huberman and his colleagues proposed several studies at the end of the 1980's and early 1990's around \emph{computational ecologies} \cite{Huberman1988}. They used the words 'Ecology' and 'ecosystems' simply to refer to a set of heterogeneous entities evolving in unpredictable environments. However the authors did not leverage ecological concepts to build and organize the agents.
Another trend  related to \emph{software ecosystems}  \cite{Messerschmitt2003} and \emph{digital ecosystems} \cite{Briscoe2006} started in the 2000's. The  ecological analogy was to the integration of heterogeneous entities and the business issues related to this heterogeneity. The analogy stays at the level of the ideas carried by the words, \emph{i.e.}, open community, decentralized development and heterogeneity, but not at the level of specific ecological scientific concepts and methods.
While some papers already leverage the deep links between Ecology and software engineering (e.g. \cite{Posnett2011}), none of them has identified the synergy as a first-class concept: \emph{``Ecology-Inspired Software Engineering''}. We think that a genuine adaptation of ecological concepts can provide innovative perspectives to build and understand software intensive systems.

\section{From Ecological to Software Concepts}
\label{sec:construction}
Biodiversity and trophic webs are two tightly coupled ecological concepts, which are found in most current ecological theories. The translations of these concepts  into the software world  can provide us with solutions to the challenge of building open yet stable large scale software systems.

\subsection{Biodiversity}

Ecologists unanimously acknowledge that a loss of \emph{biodiversity} increases the vulnerability of the system in front of changes in the environment. 
There are many types of diversity:
genetic diversity (individuals of the same species have different genotypes), functional diversity (i.e., photosynthesizer, nitrogen fixer, etc.), species diversity, spatio-temporal diversity, etc.
Ecologists keep identifying or clarifying kinds of diversity \cite{McCann2000,Orians1975} and measuring them to deepen our comprehension of the stability of ecosystems \cite{McCann2000}. The different metrics depend on the type of diversity as well as on the scale by which it is measured. 


We are convinced that injecting diversity into software can result in more adaptable and stable systems, as diversity in ecosystems yields stability. 
There is already some work on this topic (see Section \ref{sec:software-diversity}) but we envision \emph{the automated  synthesis of more kinds of diversity in software}.
 
At the software-module level, genetic diversity can be simulated by automated program mutation (minor changes in the code or in the structure of programs). Functional diversity can be mapped to the automated generation of different configurations of software modules, offering different levels of quality of service, leveraging derivation techniques from the software product line engineering.  At the global system level, we imagine an automated spatial diversity in the topology of software intensive systems that have a geographical dimension (\emph{e.g.}, sensor networks). Temporal diversity may be achieved through the provision of different software modules and connections that are activated at different points in time.


\subsection{Ecological Networks}

\emph{Ecological networks} capture different forms of direct or indirect interactions (edges) between species or populations (nodes) \cite{Montoya2006} and represent an essential structure to explain ecosystems' dynamics, evolution and robustness. 
Trophic webs (also known as food webs or food chains) is a fundamental concept  to describe species in an ecosystem. Originally proposed by Lindeman \cite{Lindeman1942}, this global model captures the different species present in an ecosystem, as well as the trophic flows (who eats who) between species. These webs are organized according to \emph{trophic levels}, which indicate the level of a given species in a food chain. Each trophic level also corresponds to a family of functionally consistent species. Then, inter trophic level relationships model prey-predator relations. What is remarkable in ecological networks in general and trophic webs in particular is their ability to remain stable in the face of perturbationss.

In software systems, many kinds of networks exist.
For instance, the prey-predator relationships can be mapped on to producer-consumer relationships in software modules. The ecological symbiotic relationships may have software counterparts in terms of library cross-dependencies. 
\emph{We believe that creating software-based networks that have similar topologies to ecological networks will improve resilience and stability capabilities.}

Finally, biodiversity and ecological networks are not orthogonal: certain types of biodiversity directly refer to the kind and number of nodes of the network (or edges as well).
In software systems, this may be translated as introducing diversity at specific nodes or edges of software networks. 
In general, we think that the software topology could drive some automated monitoring and reasoning to establish when and why to synthesize diversity in the system. 

\subsection{Software diversity}
\label{sec:software-diversity}

Software diversity has to be an essential research area for ecology-inspired software engineering. We identify two main kinds of software diversity in current litterature (and briefly mention related papers), they provide a starting point for future work on the synthesis and maintenance of diversity in software intensive systems.

\textbf{Managed software diversity} include approaches that encourage and control software diversity.
This kind of diversity is principally embodied in the work on multi-version systems (encouraging diversity) and software product lines (controlling diversity).
The N-version approach \cite{avizienis1985n} consists of implementing the same requirements N-times by N different development teams to increase reliability.
Software product line research studies how to develop similar yet diverse software products in a systematic manner \cite{pohl2005software}.

\textbf{Automated software diversity} consists of techniques to artificially introduce diversity in software. There is much work in this direction in the field of software security, for instance Kc et al. introduced the concept of instruction-set randomization to prevent code injection \cite{Kc2003}. 
However, there are also attempts to use automated software diversity for fixing bugs \cite{Goues2012} or optimize quality-of-service properties at runtime \cite{Sidiroglou-Douskos2011}.


\section{``Refuge'' Software as Reservoir of Diversity}
\label{sec:empirical}
 

The idea of \emph{refuge} in Ecology refers to conditions in which species are protected from certain threats, mostly predation threats. The simplest kind of refuge is spatial or geographic, 
for instance, the rainforest canopy is a refuge for certain species with respect to ground predators.
Berryman and Hawkins showed \cite{Berryman_Hawkins_2006} that the concept of refuge is \emph{``one of the integrating concepts in Ecology and evolution''} and showed that it is polymorphic, with an explanatory power  ranging from population dynamics to evolution. 
Species in refuges may become dominant species in the future in response to environmental changes.

\emph{A key characteristic of refuges is that they are a reservoir of diversity since they provide a means to sustain species that are not the fittest at some point in time.}

In the following, we show how the concept of \emph{refuge} in Ecology supports ecology-inspired explanations of software phenomena.

\subsection{Refuge and  Open-source Software}

A refuge refers to species and an agent of mortality \cite{Berryman_Hawkins_2006}.
In this exploratory case study, we equate ``species'' with open-source software packages, and we consider  the following agents of mortality for a project : 
1) the source or binary code is not available anymore 
2) nobody is able to understand and maintain the codebase
3) the users stop using the software. 


According to this definition, open-source projects with few users or developers live in refuges, and the  functionality they provide enables the project to survive and prevents them from joining the immense group of "dead" open source projects.
However, a key characteristic of ecological refuges refers to the ability to keep potential alternatives for biodiversity and evolution.
In order to  validate \emph{refuge} as a relevant explanatory concept for software, we  would like to observe projects that are lively but not successful, and that did seed another project that is itself successful.


\emph{Hypothesis: the ecological concept of refuge is relevant for open-source software if one can find live but not very successful open-source projects (the refuge) and whose forking descendance contains a successful project.}

In order to test this hypothesis, we need a measure: 1) that is collectible and 2) that reflects the ability of the project to survive with respect to the agents of mortality aforementioned.
We propose to measure the success of an open-source software project by its  number of forks. A fork is a kind of clone of a code base with bug fixes or additional functionalities.

This measure has the following characteristics:
1) The number of forks is proportional to the number of maintainers familiar with the code base. Hence, when they are more forks, they are more maintainers who are able to evolve the software in face of change.
2) The number of forks is proportional to the number of available functionalities and their variations, which increases the likelihood of satisfying many different users.
3) The number of forks is proportional to the number of places on the Internet where source code is available. This results in a better robustness in face of technical catastrophes (server crash) or human catastrophes (malicious behavior).
To us, these three points capture important characteristics of open-source survivability.

The number of forks also fits the perception of development dynamics in open source communities. First, the new distributed version control systems (such as Git or Mercurial) actually consider each copy of the code base as a fork and second, an open-source repository called GitHub encourages forking as a built-in feature of their infrastructure. For GitHub, forking means liberating creativity.
Since GitHub provides public access to their forking data, our success measure for open-source software is collectible.

Furthermore, the fork data  has a time dimension which also exists in ecological refuges (one refuge species may become a dominant species as the result of environmental changes). 

We crawled  the forking data of the 48 most forked projects on GitHub on Nov 15, 2011  in order to gather experimental data to validate our hypothesis (The page \url{https://github.com/popular/forked} lists 50 projects but there is one fake project and one duplicate in the list). The most forked projects on GitHub are ranked by the number of forks in the fork tree. 
Table \ref{table:stats} presents descriptive statistics on this data. In all, we crawled 36746 open-source projects from GitHub. Fork trees on GitHub are very flat since the depth of the fork tree lies between 2 and 4 and the median of the fork width is 0 (meaning that most forks have no children).

\begin{table}
\rowcolors[]{2}{gray!10}{}
\begin{tabularx}{\columnwidth}{X|c}
                        \#fork trees &                                    48 \\
                             \#forks &                                 36746 \\
     Min/Median/Max \#forks per tree &                          417/585/3030 \\
      Min/Median/Max \#forking depth &                                 2/3/4 \\
      Min/Median/Max \#forking width &                              0/0/2952 \\\end{tabularx}
\caption{Descriptive statistics on the forking data of the most forked open-source projects on GitHub.}
\label{table:stats}
\end{table}

\subsection{Are There Refuge Software Packages?}

According to the definition of software refuge and our dataset, \textbf{we observed 3 occurrences of the refuge effect}, shown in Table~\ref{table:results}.
Projects Janus, Sinatra and Delayed\_job have the desired characteristics\footnote{http://github.com/\{carlhuda/janus,sinatra/sinatra,collectiveidea/delayed\_job\}}: they are very successful in terms of forks and are themselves forks from another much less successful project.
For instance, on Nov 15, project Janus had 533 fork children while the parent fork\footnote{\url{http://github.com/carllerche/vimrc}} only had 5 forks.
In other terms, the parent fork has been at a point in time a refuge from which project Janus has emerged.
These results are a first empirical validation of the ability the ecological concept of \emph{refuge} to comprehend the nature of open-source software.

\emph{The parent projects of Janus, Sinatra and Delayed\_job acted at some point in time as reservoir of software diversity.}


Now, there are threats to this conclusion.
It may be inadequate to use the number of forks as a success measure (construct validity). To understand the extent of this threat,  we also collected the number of watchers for the three refuge projects that we observed (and their descendants). GitHub watchers declare their interest to the project, similarly to RSS feed subscribers or Facebook likes.
It turns out that the 3 refuge projects have a small number of watchers while their successful descendance (\emph{i.e.} direct or transitive succesful forks) has lots of watchers. This consolidates our intuition and argumentation on the fact that the number of forks is an appropriate success measure.
Second,  our data selection is the biggest threat to the external validity:
we only have forking data from GitHub and  we only look for refuges in the most forked projects. As we observe the refuge effect in this particular dataset, the occurrence of the effect may be purely accidental.
As we present emerging results, we propose to let subsequent work strengthen this finding.

We claimed in the introduction that ``Ecology provides a solid and rich source of inspiration for inventing new software construction paradigms''.
To us, the evidence that the refuge effect exists in open-source software shows that open-source software infrastructure should take this fact into account. 
First, organization and individuals involved in open-source software should  publish all pieces of software, even if they can not maintain or support it. This contributes to a kind of global software reservoir and the published project may be a refuge for subsequent successful code.
Second, platforms providing built-in forking capabilities and encouraging this practice facilitate the software diversity and potential refuges; GitHub is a remarkable first step in this direction.
In Ecology, similar operational decisions are made, for instance to create national parks to protect species.

\begin{table}
\rowcolors[]{2}{gray!10}{}
\begin{tabularx}{\columnwidth}{X|c}
\# Occurrences of the refuge effect &  3\\
Project Delayed\_job  \newline (\#direct forks / \#direct forks of the refuge parent project)&   315/181 \\
Project Sinatra  \newline  (\#direct forks / \#direct forks of the refuge parent project)&  \newline 341/66  \\
Project Janus \newline  (\#direct forks / \#direct forks of the refuge parent project)&  \newline 533/5\\
\end{tabularx}
\caption{Occurrences of the refuge effect in the most forked open-source projects on GitHub.}
\label{table:results}
\end{table}

\section{Conclusion}
\label{sec:conclusion}

``Ecology-inspired Software Engineering'' is a new kind of software engineering, which emphasizes ecology as a foundation for new software engineering methods.  
To us, the structure and properties of ecological systems can inspire novel software construction paradigms that increase both openness and stability.
Rather than inventing a concept, we believe that this paper names a research and engineering approach that seems to be emergent in different groups around the world.

\bibliographystyle{plain} 
\bibliography{u4f422854d620d}





\end{document}